\documentclass[a4paper,notitlepage,12pt]{article}

\begin{document}
\title{Terminal thermocapillary migration of a droplet at small Reynolds numbers and large Marangoni numbers}
\author{ Zuo-Bing Wu$^{1,2}$\footnotemark[1]\\
 $^1$State Key Laboratory of Nonlinear Mechanics, \\
 Institute of Mechanics,\\
  Chinese Academy of Sciences, Beijing 100190, China\\
  $^2$School of Engineering Science, \\University of Chinese Academy of Sciences,
  Beijing 100049, China}
 \maketitle

\footnotetext[1]{Corresponding author. Tel:. +86-10-82543955;
fax.: +86-10-82543977. \\
Email addresses: wuzb@lnm.imech.ac.cn (Z.-B. Wu)\\
Acta Mech. (2017) (in press)}

\newpage
\begin{abstract}
In this paper, the overall steady-state momentum and energy balances in the
thermocapillary migration of a droplet at small Reynolds numbers and large Marangoni numbers
are investigated to confirm the quasi-steady state assumption of the system.
The droplet is assumed to have a slight axisymmetric deformation from a sphere shape.
It is shown that under the quasi-steady state assumption,
the total momentum of the thermocapillary droplet migration system at small Reynolds numbers
is conservative.
The general solution of the steady momentum equations
can be determined with its parameters depending on the temperature fields.
However, a nonconservative integral thermal flux across the
interface for the steady thermocapillary migration of the droplet
at small Reynolds numbers and large Marangoni numbers is identified.
The nonconservative integral thermal flux
indicates that no solutions of the temperature fields exist for the steady energy equations.
The terminal thermocapillary migration of the droplet
at small Reynolds numbers and large Marangoni numbers cannot reach
a steady state and is thus in an unsteady process.

%\textbf{PACS} \  47.55.D-; 47.55.dm; 47.55.nb; 47.20.Dr\\

\end{abstract}

\newpage
\section{Introduction}

In the microgravity environment, a droplet or a bubble
placed in a nonisothermal mother liquid migrates in the direction
of increasing temperature as a result of surface tension
on the interface. This phenomenon is termed
the thermocapillary migration of the droplet or the
bubble in the fundamental hydrodynamics\cite{1}. Such nonisothermal interfacial flows
are very important in a great variety of both macroscopic and microscopic
chemical and biological engineering applications\cite{1a,1b,1c,1d,1e}.
The pioneering study in this area was carried out by Young et al(1959)\cite{2},
who gave an analytical prediction of the bubble
migration speed in the limit case of zero Reynolds(Re) number and zero
Marangoni(Ma) number. Then, Subramanian(1981)\cite{3} introduced
a quasi-steady state assumption and extended the YGB results to small Ma numbers.
Under the assumption of quasi-steady state, a series of results of theoretical analyses\cite{5,6}
and numerical simulations\cite{7,8} for migration speeds of bubbles were obtained and
are in agreement with those of experimental investigations\cite{9},
independent of Ma numbers in the systems.

Although the thermocapillary bubble migration is very well
understood, the thermocapillary droplet
migration remains a research focus due to its rather complicated behaviors with respect to the transfer of
momentum and energy through the interface of two-phase fluids.
More attention was paid to the thermocapillary droplet migration in cases of
large Re numbers and large Ma numbers. On the one hand,
the theoretical analyses and numerical simulations indicate that in the cases of
large Re numbers and large Ma numbers, the migration speed of a droplet increases with
the increase of Ma number as was
reported by Balasubramaniam \& Subramanian(2000)\cite{13} and by Ma et al(1999)\cite{14},
which is qualitatively different from the corresponding
experimental observations\cite{9,12}.
Both the theoretical
analyses and numerical simulations are based on the assumptions of
the quasi-steady state and the non-deformation of the droplet.
On the other hand, Wu \& Hu(2007) pointed out
an ill-posedness thermal flux boundary condition
for the steady thermocapillary droplet migration
 at large Re numbers and large Ma numbers\cite{13a}.
Herrmann et al(2008)\cite{15}, Wu \& Hu(2012)\cite{16} and Samereh et al(2014)\cite{30} investigated the
thermocapillary motion of deformable and non-deformable droplets by the numerical methods and
indicated that the assumption of the quasi-steady state was not valid
in cases of large Re numbers and large Ma numbers. In particular, it was shown\cite{16,30} that
the rise velocities of the droplet
against the time/the vertical position in the numerical simulations
are in qualitative agreement with those in the experimental investigations\cite{9,12}.
Yin et al(2012)\cite{30a} reported a dramatic increase of the computed stable droplet velocities when $Ma>200$
in numerical simulations, which also supports the above investigations.
These phenomena were confirmed by Wu \& Hu(2012, 2013)\cite{16,20} and they are related with the fact
that a nonconservative integral flux appears across
the interface in the steady thermocapillary droplet migration in cases of large Re numbers and large Ma numbers.
To avoid the nonconservative integral flux across the interface,
it was suggested that a thermal source is added inside the droplet. Under the
assumption of the quasi-steady state, an analytical result
for the steady thermocapillary migration of the droplet with the source
in cases of large Re numbers and large Ma numbers was obtained\cite{21}.
However, the physical mechanism of the thermocapillary droplet
migration in case of large Ma numbers is still not very well understood.

In this paper, we focus on the thermocapillary migration of a droplet
 at small Re numbers and large Ma numbers. It is closely related to the heat transfer
 in heavy oil with the Prandtl number $Pr \sim O(10^3)$ and the transport
 processes involving mass transfer\cite{23,24}.
 In general, the overall steady-state momentum and energy balances of two phases in a flow domain require that
the changes of the momentum and the energy of the domain are equal to the differences between the total
momentum and energy
entering the domain and those leaving the domain\cite{25}.
Firstly, we confirm the overall steady-state momentum balance of
the continuous phase and the fluid in the droplet
in the thermocapillary droplet migration in the co-moving frame
of reference at small Re numbers.
Then, by using the asymptotic
expansion method, we investigate the continuity of the integral
thermal flux across the interface based on the overall steady-state
energy balance in the flow domain, and analyze the
quasi-steady thermocapillary migration of the droplet in cases of small Re numbers and large Ma
numbers. Finally, the qualitative difference between the terminal thermocapillary migration of a droplet
at small Ma numbers and that in case of large Ma numbers is discussed, focusing on its intrinsic physical mechanisms.

\section{Overall steady-state momentum balance}

Consider the thermocapillary migration of a spherical droplet of
radius $R_0$, density $\gamma \rho$, dynamic viscosity $\alpha
\mu$, thermal conductivity $\beta k$, and thermal diffusivity
$\lambda \kappa$ in a continuous phase fluid of infinite extent
with density $\rho$, dynamic viscosity $\mu$, thermal conductivity
$k$, and thermal diffusivity $\kappa$ under a uniform temperature
gradient $G$ in cases of small Re numbers and large Ma numbers. The change rate of the interfacial tension between
the droplet and the continuous phase fluid as a function of temperature is
denoted by $\sigma_T(<0)$.
Axisymmetric momentum equations for the
continuous phase and for the fluid in the droplet in a laboratory cylindrical
coordinate system $({\bar \chi}, {\bar z})$ denoted by a bar are written as follows
\begin{equation}
\begin{array}{l}
\rho\frac{\partial{\bar{\bf v}}}{\partial t} + \rho \bar{\bf v} \bar{\nabla} \bar{\bf v}=
-\bar{\nabla}{\bar p} +\mu \bar{\Delta} \bar{\bf v},\\
\gamma\rho\frac{\partial{\bar{\bf v}'}}{\partial t} + \gamma\rho \bar{\bf v}' \bar{\nabla} \bar{\bf v}'=
-\bar{\nabla}{\bar p}' +\alpha\mu \bar{\Delta} \bar{\bf v}' ,
\end{array}
\end{equation}
where $\bar{\bf v}$ and $\bar{p}$ are the velocity and the pressure,
and a prime denotes quantities inside the droplet.
The solutions of
Eqs. (1) have to satisfy the boundary conditions at infinity
\begin{equation}
\bar{\bf v} \rightarrow 0, \ \ \ {\bar p} \rightarrow p_\infty,
\end{equation}
 at the center of the droplet
\begin{equation}
{\bar p}' =p'_0+ P'_0(t)
\end{equation}
and at the interface $({\bar \chi}_b,{\bar
z}_b)$ of the two-phase fluids
\begin{equation}
\begin{array}{l}
\bar {\bf v}({\bar \chi}_b,{\bar z}_b,t) =\bar {\bf v'}({\bar \chi}_b,{\bar z}_b,t),\\
{\bf n} \cdot {\bar {\bf \Pi}} - {\bf n} \cdot {\bar {\bf \Pi'}}= \sigma{\bf n}(\bar{\nabla}_s \cdot {\bf n}) - \bar{\nabla}_s \sigma,
\end{array}
\end{equation}
where $p_\infty$ is the undisturbed pressure of the continuous phase, $\bar{\bf \Pi}$ and $\bar{\bf \Pi}'$ are the stress tensors defined in terms of the local fluid velocity and the pressure fields
of the two-phase fluids.
${\bf n}$ and ${\bf \tau}$ are the unit vectors normal to the interface and tangent to its meridian, respectively.
$\bar{\nabla}_s(=\bar{\nabla} - {\bf n} \frac{\bar {\partial}}{\bar{\partial} n} )$ is the surface gradient operator.
$\sigma[=\sigma_0 +\sigma_T ({\bar T}-T_0)]$ is the coefficient of the interface tension decreasing linearly with the temperature ${\bar T}$.

Under the quasi-steady state assumption, the velocity
 fields are fully established at every moment of time.
After an initial unstable migration
process, the droplet migration may reach a steady state,
i.e., migrating
with a constant speed $V_{\infty}$.
In this case, the momentum equations (1) with the boundary conditions (2)(3)(4) can be established
in a coordinate system moving with the droplet velocity $V_\infty$
under the coordinate transformation
\begin{equation}
\bar{\bf r} = {\bf r} + V_\infty t {\bf k}
\end{equation}
and variable transformations
\begin{equation}
\begin{array}{lll}
\bar{\bf v}(\bar{\bf r},t) = {\bf v}({\bf r}) + V_\infty {\bf k}, & \bar{p} = p+ p_\infty,
& \bar{T}(\bar{\bf r},t) = T({\bf r}) +T_0 + GV_\infty t,\\
 \bar{\bf v'}(\bar{\bf r},t) = {\bf v'}({\bf r}) + V_\infty {\bf k}, & \bar{p}' = p'+ P'_0(t),
& \bar{T'}(\bar{\bf r},t) = T'({\bf r}) +T_0 + GV_\infty t .
\end{array}
\end{equation}
The droplet is assumed to have a slight axisymmetric deformation from its original sphere shape.
The radius of the droplet $R_0$, the velocity $v_0=-\sigma_T G
R_0/\mu$ and $GR_0$ are used as the reference quantities to make the
coordinates, the velocity and the temperature dimensionless. The momentum
equations (1) combined with the continuous equations can be
rewritten in the following dimensionless form in a spherical
coordinate system ($r,\theta$) moving with the droplet velocity $V_{\infty}$
\begin{equation}
\begin{array}{l}
\nabla \cdot {\bf vv}= \nabla \cdot {\bf \Pi},\\
\gamma \nabla \cdot {\bf v'v'}= \nabla \cdot {\bf \Pi'},
\end{array}
\end{equation}
where ${\bf v}=(v_r,v_\theta)$,
${\bf \Pi}= -p {\bf I} + \frac{1}{Re} [\nabla {\bf v} +(\nabla {\bf v})^T]$ and
${\bf \Pi'}= -p' {\bf I} + \frac{\alpha}{Re} [\nabla {\bf v'} +(\nabla {\bf v'})^T]$.
The Reynolds number is defined as
\begin{equation}
Re=\frac{\rho v_0 R_0}{\mu}.
\end{equation}
The boundary conditions (2)(3)(4) are rewritten in the dimensionless form as
\begin{equation}
(v_r, v_\theta) \to (-V_\infty \cos \theta, V_\infty \sin \theta), \ \ \
p \to 0
\end{equation}
 at infinity,
\begin{equation}
p' = p'_0
\end{equation}
 at the center of the droplet
 and
\begin{equation}
\begin{array}{l}
v_n(R,\theta) =v'_n(R,\theta)=0,\\
v_\tau(R,\theta) =v'_\tau(R,\theta),\\
{\bf n} \cdot {\bf \Pi} - {\bf n} \cdot {\bf \Pi'}=\frac{1}{We} [\sigma^* {\bf n} (\nabla_s \cdot {\bf n})- \nabla_s \sigma^*]
\end{array}
\end{equation}
at the interface $R[=1+  f_1(\theta)Re + f_2(\theta)Re^2 + O(Re^3)]$ of the two-phase fluids.
The Capillary and Weber numbers are defined, respectively, as
\begin{equation}
Ca=\frac{v_0 \mu}{\sigma_0}
\end{equation}
and
\begin{equation}
We=ReCa=\frac{\rho v_0^2 R_0}{\sigma_0}.
\end{equation}
The non-dimensional surface tension coefficient is now expressed as $\sigma=1-Ca (T+V_\infty t)=\sigma^* -CaV_\infty t$.
$P'_0(t)=p_\infty+\frac{1}{Re}V_\infty t \nabla_s \cdot {\bf n} \approx p_\infty+\frac{2}{Re}V_\infty t$.
Thus, once the droplet migration
reaches a steady state, the above problem (1)-(4) in the laboratory
coordinate system can be described by the steady momentum equations
(7) with the boundary conditions (9)(10)(11) in the coordinate
system moving with the droplet velocity. The overall steady-state momentum balance of the two phases
in the flow domain in the co-moving frame of reference is also guaranteed.

To confirm the overall steady-state momentum balance of the two phases under the above-mentioned assumption,
the entire system $V$ is divided into two subsystems, i.e., the continuous phase domain $W$ $(r\in
[R,r_{\infty}],\theta\in[0,\pi])$ and the droplet region $W'$
$(r\in [0,R],\theta\in[0,\pi])$. We can write the time rate of the momentum per unit volume ${\bf \Xi}$ in the system $V$ as
\begin{equation}
\frac{d}{dt} \int_V {\bf {\Xi}} dV = \frac{d}{dt} \int_{W} {\bf v} dV + \gamma \frac{d}{dt} \int_{W'}  {\bf v'} dV.
\end{equation}
For the righthand side terms in Eq. (14), we apply the generalized transport theorem\cite{25} and the momentum equations (7)
\begin{equation}
\begin{array}{l}
\frac{d}{dt} \int_{W} {\bf v} dV=
\int_{W} \nabla \cdot {\bf vv} dV =
\int_{S_\infty} {\bf n} \cdot {\bf \Pi} dS - \int_{S_1} {\bf n} \cdot {\bf \Pi} dS,\\
\gamma\frac{d}{dt} \int_{W'} {\bf v'} dV=
\gamma\int_{W'} \nabla \cdot {\bf v'v'} dV =
\int_{S_1} {\bf n} \cdot {\bf \Pi'}  dS,
\end{array}
\end{equation}
where $S_1$ is the interface of the two-phase fluids.
Substituting these expressions into Eq. (14), we have
\begin{equation}
\begin{array}{lll}
\frac{d}{dt} \int_V {\bf {\Xi}} dV &=& \int_{S_\infty} {\bf n}\cdot {\bf \Pi} dS - \int_{S_1} ({\bf n} \cdot {\bf \Pi}-{\bf n} \cdot{\bf \Pi'}) dS.
\end{array}
\end{equation}
For the axisymmetric droplet, the first term on the righthand side of Eq. (16) is rewritten as
\begin{equation}
\begin{array}{ll}
\int_{S_\infty} {\bf n}\cdot {\bf \Pi} dS &= \int_{S_\infty} \Pi_{rr} dS {\bf i_r} + \int_{S_\infty} \Pi_{r\theta} dS {\bf i_{\theta}}\\
&= \int_{S_\infty} (\Pi_{rr} \sin \theta +\Pi_{r\theta} \cos \theta) dS {\bf i_\chi} + \int_{S_\infty} (\Pi_{rr} \cos \theta -\Pi_{r\theta} \sin \theta) dS {\bf i_z},
\end{array}
\end{equation}
where $\Pi_{rr}=-p+\frac{2}{Re}\frac{\partial v_r}{\partial r}$ and $\Pi_{r\theta}=\frac{1}{Re}(\frac{1}{r}\frac{\partial v_r}{\partial \theta} + \frac{\partial v_\theta}{\partial r} - \frac{v_\theta}{r})$.
$({\bf i_r}, {\bf i_\theta})$ and
$({\bf i_\chi}, {\bf i_z})$ are the unit vectors in the spherical coordinate system $(r, \theta)$ and the cylindrical one $(\chi, z)$, respectively.
Since Eq. (9) gives only an approximation of the
velocity and the pressure fields at infinity, we have to obtain their asymptotic
expansions with respect to the small Re number.
 In the Appendix, a general solution of the momentum equations (7) for the thermocapillary migration of the droplet
  is obtained in the form of an expansion in powers of the Re number.
 The scaled velocity and pressure fields in the continuous phase are described as
\begin{equation}
\begin{array}{ll}
v_r &= v^{(0)}_{r} + Re v^{(1)}_{r} + o(Re)\\
&=-V_\infty (1-\frac{1}{r^3})\cos \theta - Re \frac{C}{6r^2}(1-\frac{1}{r^2})(3 \cos^2 \theta -1)+ o(Re),\\
v_\theta &= v^{(0)}_{\theta} + Re v^{(1)}_{\theta}+ o(Re)\\
&=V_\infty (1+\frac{1}{2r^3})\sin \theta + Re \frac{C}{3r^4} \sin \theta \cos \theta + o(Re),\\
p &= \frac{1}{Re}p^{(0)} + p^{(1)} + o(1)\\
&= -\frac{V^2_\infty}{4r^6} +[\frac{V^2_\infty}{2}(\frac{1}{r^3} -\frac{1}{4r^6}) -\frac{C}{3r^3}] (3 \cos^2 \theta-1)+ o(1),
\end{array}
\end{equation}
where $V_\infty$ and $C$ are constants.
Using the above expressions, Eq. (17) is derived as
\begin{equation}
\begin{array}{lll}
&&\int_{S_\infty} {\bf n}\cdot {\bf \Pi} dS = \int_0^{\pi} ({\bf n}\cdot {\bf \Pi})|_{r_\infty} r^2_\infty \sin \theta d\theta\\
&=&\{-r^2_\infty \int_0^\pi p \sin^2 \theta d\theta + \frac{r^2_\infty}{Re} \int_0^\pi [2\frac{\partial v_r}{\partial r} \sin \theta
+(\frac{1}{r}\frac{\partial v_r}{\partial \theta} +\frac{\partial v_\theta}{\partial r} -\frac{v_\theta}{r}) \cos \theta] \sin \theta d \theta \} {\bf i_\chi}\\
&&-\{r^2_\infty \int_0^\pi p \cos \theta \sin \theta d\theta - \frac{r^2_\infty}{Re} \int_0^\pi [2\frac{\partial v_r}{\partial r} \cos \theta
-(\frac{1}{r}\frac{\partial v_r}{\partial \theta} +\frac{\partial v_\theta}{\partial r} -\frac{v_\theta}{r}) \sin \theta] \sin \theta d \theta \} {\bf i_z}\\
& =& \{ -r^2_\infty \int_0^\pi p^{(1)} \sin^2 \theta d\theta \\
&&+ r^2_\infty \int_0^\pi [2\frac{\partial v^{(1)}_{r}}{\partial r} \sin \theta
+(\frac{1}{r}\frac{\partial v^{(1)}_{r}}{\partial \theta} +\frac{\partial v^{(1)}_{\theta}}{\partial r} -\frac{v^{(1)}_{\theta}}{r}) \cos \theta] \sin \theta d \theta +o(1) \} {\bf i_\chi} \\
&&-\{r^2_\infty \int_0^\pi p^{(1)}_1 \cos \theta \sin \theta d\theta \\
&&- r^2_\infty \int_0^\pi [2\frac{\partial v^{(1)}_{r}}{\partial r} \cos \theta
-(\frac{1}{r}\frac{\partial v^{(1)}_{r}}{\partial \theta} +\frac{\partial v^{(1)}_{\theta}}{\partial r} -\frac{v^{(1)}_{\theta}}{r}) \sin \theta] \sin \theta d \theta +o(1) \} {\bf i_z} \\
&= &[\frac{\pi}{2r_\infty} (\frac{V^2_\infty}{8} + \frac{C}{r^2_\infty} +\frac{31V^2_\infty}{32r^3_\infty}) + o(1)] {\bf i_\chi} +  o(1) {\bf i_z} \\
& \sim &\{O(r^{-1}_\infty) + o(1) \} {\bf i_\chi} +  o(1) {\bf i_z} \\
& \to&0.
\end{array}
\end{equation}
Moreover, the second term in the righthand side of Eq. (16) is simplified by means of the boundary conditions (11) and the Stokes theorem as
\begin{equation}
\begin{array}{l}
\int_{S_1} ({\bf n} \cdot {\bf \Pi}-{\bf n} \cdot{\bf \Pi'})   dS\\
= \frac{1}{We}  \int_{S_1}[\sigma^* {\bf n} (\nabla_s \cdot {\bf n}) - \nabla_s \sigma^* ] dS\\
=-\frac{1}{We}  \oint_{C^-} \sigma^*  dl {\bf \tau}+\frac{1}{We}  \oint_{C^+} \sigma^*  dl {\bf \tau}\\
= 0,
\end{array}
\end{equation}
where $l$ is the arc length along the closed contour  $C^-/C^+$ (the left/right limit of the equator $C$)
on the interface $S_1$.
Finally, substituting the expressions (19)(20) into Eq.(16), we have
\begin{equation}
\frac{d}{dt} \int_V {\bf \Xi} dV =0.
\end{equation}
This implies the overall steady-state momentum balance
of two phases in the flow domain in the co-moving frame of reference.
Therefore, under the quasi-steady state assumption, the total momentum of the thermocapillary droplet migration system in case of small Re numbers
is conservative.

The quasi-steady state assumption not only implies the existence of the steady velocity fields for Eq. (7),
but also requires that the net force acting on the droplet by the continuous phase
must vanish, i.e., $\int_{S_1} {\bf n} \cdot {\bf \Pi} dS =0$. Substituting it into Eq. (20), we have $\int_{S_1} {\bf n} \cdot {\bf \Pi'} dS =0$.
In this case, using the above result and the expression (19), Eq. (15) is reduced to  $\frac{d}{dt} \int_W {\bf v} dV =\gamma \frac{d}{dt} \int_{W'} {\bf v'} dV =0$.
Thus, under the quasi-steady state assumption,
the momentums in two subsystems of the thermocapillary droplet migration system in the case of small Re numbers are also conservative.

Note that the velocity and the pressure fields at infinity at large Re numbers\cite{29},
obtained by  truncating the expressions (18) at $O(Re)$ and $O(1)$, respectively,
can be taken as a special case of small Re numbers.
So, in view of the above derivations, the total momentum of the steady thermocapillary droplet migration system at large Re numbers
and the momentums in its two subsystems
are also conservative\cite{20}.

It is meaningful to investigate whether the overall steady-state momentum balance in the
thermocapillary droplet migration  is valid when the droplet is assumed to be non-deformable and kept in the sphere shape($R=1$).
In this case, the stress balance condition at the interface in Eqs. (11) is simplified as
\begin{equation}
{\bf n} \cdot {\bf \Pi} - {\bf n} \cdot {\bf \Pi'}=\frac{1}{We} (2 \sigma^* {\bf n}- \frac{\partial \sigma^*}{\partial \theta} {\bf \tau}),
\end{equation}
where $\nabla_s \cdot {\bf n}= \nabla \cdot {\bf n} = 2$.
Using the above boundary condition, the second term in the righthand side of Eq. (16) becomes
\begin{equation}
\begin{array}{lll}
\int_{S_1} ({\bf n} \cdot {\bf \Pi}-{\bf n} \cdot{\bf \Pi'})   dS&=& \frac{1}{We}  \int_{S_1}(2\sigma^* {\bf n} - \frac{\partial \sigma^*}{\partial \theta} {\bf \tau})  dS\\
&=& \frac{1}{We}  \int_{S_1} (2\sigma^* \sin \theta - \frac{\partial \sigma^*}{\partial \theta} \cos \theta ) dS {\bf i_\chi} \\
&&+ \frac{1}{We} \int_{S_1} (2\sigma^* \cos \theta + \frac{\partial \sigma^*}{\partial \theta} \sin \theta )  dS {\bf i_z}\\
&=& \frac{1}{We} \int_0^\pi \sigma^* d \theta {\bf i_\chi}.
\end{array}
\end{equation}
Due to the axisymmetric features of the interface $S_1$, the interaction $\frac{1}{We} \int_0^\pi \sigma^* d \theta {\bf i_\chi}$
in the cross section (${\bf i_\chi}, {\bf i_z}$) will be balanced with that
in the cross section ($-{\bf i_\chi}, {\bf i_z}$).
Taking into account of the balance of the interactions in the axisymmetric interface, we have
$\int_{S_1} ({\bf n} \cdot {\bf \Pi}-{\bf n} \cdot{\bf \Pi'}) dS$ =0.
Thus, substituting the above result and the expression (19) into Eq.(16), we have
\begin{equation}
\frac{d}{dt} \int_V {\bf \Xi} dV =0.
\end{equation}
This implies that, under the assumptions of the quasi-steady state and the non-deformation of the droplet,
the total momentum of the thermocapillary droplet migration system
is also conservative.

\section{Overall steady-state energy unbalance}
The axisymmetric energy equations for the
continuous phase and for the fluid in the droplet in a laboratory
cylindrical coordinate system $({\bar \chi}, {\bar z})$ denoted by a bar are written as
\begin{equation}
\begin{array}{l}
\frac{\partial{\bar{T}}}{\partial t} + \bar{\bf v} \bar{\nabla} \bar{T}= \kappa \bar{\Delta} \bar{T},\\
\frac{\partial{\bar{T'}}}{\partial t} + \bar{\bf v'} \bar{\nabla} \bar{T'}= \lambda \kappa \bar{\Delta} \bar{T'},
\end{array}
\end{equation}
where $\bar{T}$ is the temperature.
The solutions of
Eqs. (25) have to satisfy the boundary condition at infinity
\begin{equation}
{\bar T} \rightarrow {\bar T}_0 + G{\bar z},
\end{equation}
where ${\bar T}_0$ is the undisturbed temperature of the continuous phase
and the boundary conditions at the interface $({\bar \chi}_b,{\bar
z}_b)$ of the two-phase fluids
\begin{equation}
\begin{array}{l}
\bar T({\bar \chi}_b,{\bar z}_b,t) =\bar T'({\bar \chi}_b,{\bar z}_b,t),\\
\frac{\partial{\bar T}}{\partial n}({\bar \chi}_b,{\bar z}_b,t)= \beta
\frac{\partial{\bar T'}}{\partial n}({\bar \chi}_b,{\bar z}_b,t).
\end{array}
\end{equation}
Under the quasi-steady state assumption,
the temperature fields are fully established
at every moment of time.
In a similar way, under the coordinate and variable transformations (5)(6),
the energy equations (25) combined with the continuous equations can be
rewritten in the following dimensionless form in the spherical
coordinate system ($r,\theta$) moving with the droplet velocity $V_{\infty}$
\begin{equation}
\begin{array}{l}
V_{\infty}+\nabla \cdot ({\bf v}T) = \frac{1}{Ma} \Delta T,\\
V_{\infty}+\nabla \cdot ({\bf v'}T') =\frac{\lambda}{Ma} \Delta T',
\end{array}
\end{equation}
where the Marangoni number is defined as
\begin{equation}
Ma=\frac{v_0R_0}{\kappa}.
\end{equation}
The boundary conditions (26)(27) are rewritten in the dimensionless form as
\begin{equation}
T \to r \cos \theta
\end{equation}
 at infinity and
 \begin{equation}
\begin{array}{l}
T(R,\theta) =T'(R,\theta),\\
\frac{\partial{T}}{\partial n}(R,\theta) = \beta
\frac{\partial{T'}}{\partial n}(R,\theta)
\end{array}
\end{equation}
at the interface of the two-phase fluids.
In the same way, the assumption can be confirmed by the overall steady-state energy balance of the two phases
in the flow domain in the co-moving frame of reference.
  To obtain the overall steady-state energy transport of two phases in the flow domain, Eq. (28) is integrated
in the continuous phase domain $W$ and the droplet region $W'$.
Following a similar derivations as in \cite{20}, we have
\begin{equation}
\begin{array}{ll}
\int_{S_1} \frac{\partial T}{\partial n} dS = &r_\infty^2 \int_0^{\pi} \frac{\partial
T}{\partial r}|_{r_\infty} \sin \theta d\theta - Ma r_\infty^2
\int_0^{\pi} (u T)|_{r_{\infty}} \sin \theta d\theta \\
&- \frac{2V_\infty Ma}{3}
(r_\infty^3-\frac{1}{2} \int_0^\pi R^3 \sin \theta d \theta)
\end{array}
\end{equation}
and
\begin{equation}
 \int_{S_1} \frac{\partial {T'}}{\partial n}dS
=\frac{V_\infty Ma}{3 \lambda} \int_0^\pi R^3 \sin \theta d \theta.
\end{equation}
Since Eq. (30) only gives an approximation of the
temperature field at infinity, we have to obtain an asymptotic
expansion of $T$ for the integration of Eq. (32).
In the limit case of large Ma numbers, we may apply
the perturbation theory to determine the asymptotic behavior of $T$ at $r \gg 1$.
In general, the steady migration velocity $V_\infty[=\frac{1}{2+3\alpha} \int_0^\pi T(1,\theta) \sin \theta \cos \theta d \theta>0]$
is a nonlinear function of the Ma number and has
three possible functional relationships with the Ma number in the case of the large Ma numbers.

(A) $V_\infty$ is an increasing function of the Ma number,
$V_\infty= V^{[\varphi]}_\infty G(Ma) +o[G(Ma)]$, where $V^{[\varphi]}_\infty>0$ and $O[G(Ma)]>O[1]$.
The function $G(Ma)$ is of the highest order in the nonlinear functional set to show the relationship of $V_\infty$ with Ma,
such as $Ma^{\xi}$, $Ma^{\eta} \ln Ma$, $(\ln Ma)^{\zeta}$, $\cdots$, $\xi>0$, $\eta>0$ and $\zeta>0$.
Taking $\epsilon=1/\sqrt{V_\infty Ma}$ as a small parameter, we have
$v_r=v^{[0]}_r + O[\epsilon]$, $v_\theta=v^{[0]}_{\theta} + O[\epsilon]$ and $T=T^{[0]} + O[\epsilon]$.
The energy equation for the continuous phase in Eq. (28) truncated at the zeroth order $O[1]$ may be written as
\begin{equation}
1+ v^{[0]}_r \frac{\partial T^{[0]}}{\partial r} + \frac{v^{[0]}_{\theta}}{r}\frac{\partial T^{[0]}}{\partial \theta} =0.
\end{equation}
Introducing $T^{[0]}_*=T^{[0]}-r \cos \theta$, we have
\begin{equation}
v^{[0]}_r \frac{\partial T^{[0]}_*}{\partial r} + \frac{v^{[0]}_\theta}{r}\frac{\partial T^{[0]}_*}{\partial \theta} =
 v^{[0]}_\theta \sin \theta- v^{[0]}_r \cos \theta -1 .
\end{equation}
In the case of small Re numbers, the velocity fields $(v_r, v_\theta)$ have the expansions against Re in Eqs. (18).
This leads to $v^{[0]}_r=v^{(0)}_{r} + Re v^{(1)}_r + o(Re)$ and $v^{[0]}_\theta=v^{(0)}_{\theta} + Re v^{(1)}_{\theta} + o(Re)$.
Using the coordinate transformation $\xi=r$ and $\eta=\Psi^{(0)}=-\frac{1}{2}\sin^2 \theta(r^2-1/r)$, Eq.(35) can be expressed as
\begin{equation}
\begin{array}{ll}
\frac{\partial T^{[0]}_*}{\partial \xi} &=
(v^{[0]}_\theta \sin \theta - v^{[0]}_r \cos \theta -1)/v^{[0]}_r - Re (v^{(1)}_{r} \eta_r +  \frac{v^{(1)}_{\theta}}{r} \eta_\theta)/v^{[0]}_r \frac{\partial T^{[0]}_*}{\partial \eta} \\
&\approx (v^{(0)}_{\theta} \sin \theta - v^{(0)}_{r} \cos \theta -1)/v^{(0)}_{r} + Re (v^{(1)}_{\theta} \sin \theta - v^{(1)}_{r} \cos \theta)/v^{(0)}_{r}\\
&-Re (v^{(0)}_{\theta} \sin \theta - v^{(0)}_{r} \cos \theta -1)v^{(1)}_{r}/{v^{(0)}_r}^2 -Re (v^{(1)}_{r} \eta_r +  \frac{v^{(1)}_{\theta}}{r} \eta_\theta)/v^{(0)}_{r}\frac{\partial T^{(0)}_*}{\partial \eta} + o(Re)\\
&= (v^{(0)}_{\theta} \sin \theta - v^{(0)}_{r} \cos \theta -1)/v^{(0)}_{r} +O(Re),
\end{array}
\end{equation}
where $1/v^{[0]}_r \approx 1/v^{(0)}_{r}-Re v^{(1)}_{r}/{v^{(0)}_{r}}^2 + o(Re)$.
The analytical result of the outer temperature field for the continuous phase may be obtained as
\begin{equation}
\begin{array}{ll}
T&= r \cos \theta + T^{[0]}_* + O[\epsilon] \\
&=r \cos \theta + \int^{r}_\infty (v^{(0)}_{\theta} \sin \theta -v^{(0)}_{r} \cos \theta
-1)/v^{(0)}_{r}|_{\Psi^{(0)}} d\tilde{r} + O(Re)+ O[\epsilon],
\end{array}
\end{equation}
where the operator "+" before
the integral is selected to preserve the monotonously increasing
trend of $T(r,0)$ with $r(>1)$ for the continuous phase. Using the expressions (18), we have
\begin{equation}
T=r \cos \theta + \int^{r}_\infty \frac{1}{\tilde{r}^3-1}
\frac{2-3 \sin^2 \theta}{2 \cos \theta}|_{\Psi^{(0)}} d\tilde{r} + O(Re)+ O[\epsilon].
\end{equation}
 Replacing $\theta$ by $\Psi^{(0)}$ in Eq. (38), we
have
\begin{equation}
T=r \cos \theta - \int_r^\infty \frac{1}{\tilde{r}^3-1} \frac{1 +3\Psi^{(0)} / (\tilde{r}^2-1/\tilde{r})}{\pm
\sqrt{1+2\Psi^{(0)}/(\tilde{r}^2-1/\tilde{r})}}|_{\Psi^{(0)}} d\tilde{r} + O(Re)+ O[\epsilon],
\end{equation}
where the operator "$\pm$" in the integral is to be determined by the value of
$\theta$ (the operator $"+"/"-"$ corresponds to $\theta \in
[0,\pi/2)/[\pi/2,\pi)$).
 At $r \gg 1$, the
result (39) can be expressed as
\begin{equation}
\begin{array}{ll}
T &\approx r \cos \theta - \int_r^\infty \frac{1}{\tilde{r}^3}
\frac{1+ 3 \Psi^{(0)}  / \tilde{r}^2} {\pm
\sqrt{1 +2\Psi^{(0)}/\tilde{r}^2}}|_{\Psi^{(0)}}
d\tilde{r} + O(Re) + O(\epsilon)\\
 &= r \cos \theta -\frac{1}{2r^2} \cos \theta + O(Re)+ O[\epsilon],
 \end{array}
\end{equation}
where $\Psi^{(0)} \approx -\frac{1}{2}\sin^2 \theta r^2$.
Using the above temperature field at the infinity, Eq. (32) may be simplified as
\begin{equation}
\begin{array}{ll}
\int_{S_1} \frac{\partial{T}}{\partial{n}} dS
&= -\frac{1}{3\epsilon^2}
(3-\frac{1}{r^3_{\infty}}) +  \frac{1}{3\epsilon^2} \int_0^\pi R^3 \sin \theta d\theta + O(Re)
+ O[\frac{1}{\epsilon}].\\
 \end{array}
\end{equation}
From Eq. (33) and Eq. (41), we have
\begin{equation}
\begin{array}{lll}
\int_{S_1} (\beta \frac{\partial{T'}}{\partial{n}} -
\frac{\partial{T}}{\partial{n}}) dS
 &=&\frac{1}{3\epsilon^2} [(3- \frac{1}{r_\infty^3})  + (\frac{\beta}{\lambda}-1) \int_0^{\pi} R^3 \sin \theta d\theta] + O(Re) + O[\frac{1}{\epsilon}]\\
 & \approx & \frac{1}{3} V_{\infty}Ma [(1+ \frac{2 \beta}{\lambda})   +3Re^2(\frac{\beta}{\lambda}-1) \int_0^{\pi} f_2(\theta) \sin \theta d\theta ] \\
 &&+ O[V_{\infty}Ma] O(Re^3) + O(Re) + O[({V_{\infty}Ma})^{\frac{1}{2}}]\\
& \approx & \frac{1}{3} V^{[\varphi]}_{\infty}GMa [(1+ \frac{2 \beta}{\lambda})   +3Re^2(\frac{\beta}{\lambda}-1) \int_0^{\pi} f_2(\theta) \sin \theta d\theta ] \\
 &&+o[GMa]+ O[GMa]O(Re^3) + O(Re) + O[G^{\frac{1}{2}}Ma^{\frac{1}{2}}],
 \end{array}
\end{equation}
where $R^3 = 1+3f_2(\theta)Re^2+O(Re^3)$.
Since both $\beta$ and $\lambda$ are positive, we have
\begin{equation}
\beta \int_{S_1} \frac{\partial{T'}}{\partial{n}}(R,\theta) dS - \int_{S_1}
\frac{\partial{T}}{\partial{n}}(R,\theta) dS \sim O[GMa] > O[Ma] \gg 0.
\end{equation}

(B) $V_\infty$ is a decreasing function of the Ma number or a constant,
$V_\infty=V^{[0]}_\infty + V^{[\varphi]}_\infty G(Ma)+o[G(Ma)]$, where $V^{[0]}_\infty > 0$ and $O[G(Ma)]<O[1]$.
$V^{[0]}_\infty$ is related with the zeroth order inner temperature near the interface within the thermal boundary layer, which is to be matched
 with the zeroth order outer temperature outside the thermal boundary layer\cite{6,21}.
The function $G(Ma)$ is of the lowest order in the nonlinear functional set to show the relationship of $V_\infty$ with Ma,
such as $Ma^{-\xi}$, $Ma^{-\eta} ln Ma$, $(\ln Ma)^{-\zeta}$, $\cdots$, $\xi>0$, $\eta>0$ and $\zeta>0$.
Taking $\epsilon=1/\sqrt{Ma}$ as a small parameter, we have
$v_r=v^{[0]}_r +O[\epsilon]$, $v_\theta=v^{[0]}_{\theta} + O[\epsilon]$, $T=T^{[0]} + O[\epsilon]$
and $V_\infty=V^{[0]}_\infty + V^{[\varphi]}_\infty G(1/\epsilon^2) +o[G(1/\epsilon^2)]$.
The energy equation for the continuous phase  in Eq. (28)  truncated at the zeroth order $O[1]$ may be written as
\begin{equation}
V^{[0]}_\infty+ v^{[0]}_r \frac{\partial T^{[0]}}{\partial r} + \frac{v^{[0]}_{\theta}}{r}\frac{\partial T^{[0]}}{\partial \theta} =0.
\end{equation}
In a similar manner, the outer temperature field for the continuous fluid at $r \gg 1$ can be expressed as
\begin{equation}
T = r \cos \theta -\frac{1}{2r^2} \cos \theta + O(Re) + O[\epsilon].
\end{equation}
Using the above temperature field at the infinity, Eq. (32) is simplified as
\begin{equation}
\begin{array}{ll}
\int_{S_1} \frac{\partial{T}}{\partial{n}} dS
&= -\frac{V_\infty}{3\epsilon^2}
(3-\frac{1}{r^3_{\infty}}) +  \frac{V_\infty}{3\epsilon^2} \int_0^\pi R^3 \sin \theta d\theta + O(Re)
+ O[\frac{V_\infty}{\epsilon}].
 \end{array}
\end{equation}
From Eq. (33) and Eq. (46), we have
\begin{equation}
\begin{array}{lll}
\int_{S_1} (\beta \frac{\partial{T'}}{\partial{n}} -
\frac{\partial{T}}{\partial{n}}) dS
 &=&\frac{V_\infty}{3\epsilon^2} [(3- \frac{1}{r_\infty^3})  + (\frac{\beta}{\lambda}-1) \int_0^{\pi} R^3 \sin \theta d\theta] + O(Re) + O[\frac{V_\infty}{\epsilon}]\\
 & \approx & \frac{1}{3} V^{[0]}_{\infty}Ma [(1+ \frac{2 \beta}{\lambda})   +3Re^2(\frac{\beta}{\lambda}-1) \int_0^{\pi} f_2(\theta) \sin \theta d\theta ] \\
 &&+ O[GMa] + O[Ma]O(Re^3) + O(Re) + O[Ma^\frac{1}{2}].
 \end{array}
\end{equation}
Since both $\beta$ and $\lambda$ are positive, we have
\begin{equation}
\beta \int_{S_1} \frac{\partial{T'}}{\partial{n}}(R,\theta) dS - \int_{S_1}
\frac{\partial{T}}{\partial{n}}(R,\theta) dS \sim O[Ma] \gg 0.
\end{equation}

To sum up, Eq. (43) and Eq. (48) may be rewritten as
\begin{eqnarray}
\beta \int_{S_1} \frac{\partial{T'}}{\partial{n}}(R,\theta) dS  \gg \int_{S_1}
\frac{\partial{T}}{\partial{n}}(R,\theta) dS.
\end{eqnarray}
 From Eq. (31), the integral thermal flux boundary condition across
the interface can be expressed as
\begin{eqnarray}
\beta \int_{S_1} \frac{\partial{T'}}{\partial{n}}(R,\theta) dS = \int_{S_1}
\frac{\partial{T}}{\partial{n}}(R,\theta) dS.
\end{eqnarray}
So, if the overall steady-state energy of two phases in the flow
domain is balanced, Eq. (49)
should be reduced to Eq. (50), which seems impossible. It is
termed as a nonconservative integral thermal flux across the
interface for the steady thermocapillary droplet migration at small Re numbers and large Ma
numbers. This implies the overall steady-state energy
unbalance of two phases in the flow domain in the co-moving frame
of reference and indicates that the steady temperature fields are impossible.
The terminal thermocapillary droplet migration
at small Re numbers and large Ma numbers cannot reach a steady state.
Its terminal state is thus unsteady.

\section{Conclusions and discussions}
In summary, to verify the quasi-steady state assumption in
the thermocapillary migration of a droplet in cases of small Re numbers and large Ma numbers,
we have investigated the overall steady-state momentum and energy balances of the
system.
The droplet is assumed to have a slight axisymmetric deformation from a sphere shape.
Under the quasi-steady state assumption,
the total momentum of the thermocapillary droplet migration system at small Re numbers
has been shown as conservative.
Although the general solution of the steady momentum equations can be
determined independently, its parameters still depend on the temperature fields.
However,  a nonconservative integral thermal flux across the
interface for the steady thermocapillary migration of the droplet at small Re numbers and large Ma numbers
has been identified. The nonconservative integral thermal flux
indicates that there is no solution of the temperature fields for the steady energy equations.
The terminal thermocapillary migration of the droplet
at small Re numbers and large Ma numbers cannot reach a
steady state and is thus in an unsteady process.

From the above analysis and the previous investigations of the thermocapillary droplet migration
 at large Re numbers and large Ma numbers\cite{20}, we can conclude that the terminal thermocapillary
migration of a droplet at large Ma numbers is in an unsteady state for
the system at any Re numbers. Moreover, it was demonstrated theoretically that the terminal
thermocapillary migration of a droplet at small Ma numbers is in a steady state\cite{20}.
The qualitative difference between the terminal thermocapillary migration of a droplet at small Ma numbers
and that at large Ma numbers is
due to the evolution of the mode and the intensity of the heat transfer in the system.
In general, when the droplet migrates upward, the thermal convection
and conduction are two possible ways of the heat transfer in the continuous phase fluid.
Either of them can make a redistribution of the thermal energy around the droplet.
However, the droplet can only obtain the thermal energy though the thermal conduction across the interface.
The thermal energy is not only transferred into the droplet from the top surface but also out the droplet from the bottom surface.
The thermal convection in the droplet can only make a redistribution of the thermal energy.
In case of small Ma numbers, the heat transfer in the system duo to
the thermal conduction across/around the droplet is stronger than that due to the thermal convection around the droplet,
 so that both the continuous phase fluid and
the droplet can obtain simultaneously the thermal energy.
The external and internal temperature of the droplet may have a fixed difference as shown in Eq. (6).
This leads to a steady migration process.
The steady thermocapillary migration of a droplet in case of small Ma numbers  was
verified experimentally\cite{2,32}.
At large Ma numbers, the heat transfer in the system due to the thermal convection around the droplet is stronger than
that due to the thermal conduction across/around the droplet. During the upward migration process,
the increment of the internal temperature of the droplet is far smaller than that of the external temperature of the droplet,
so that the difference between the external and internal temperatures of the droplet increases more and more,
and cannot reach a fixed value as shown in Eq. (6). This reveals an unsteady migration process.
The unsteady thermocapillary migration of a droplet at large Ma numbers was
verified experimentally\cite{9,12}.
These intrinsic physical mechanisms of the thermocapillary droplet migration at small numbers and large Ma numbers
are conducible to the understanding of the interaction of the droplets in both numerical simulations\cite{26,27}
and experimental investigations \cite{28}.

\newpage
\textbf{Acknowledgments} This research is supported by the National Natural Science Foundation
of China through the Grants No. 11172310 and No. 11472284.
The author thanks the IMECH research computing facility for
assisting in the computation.

\newpage
\textbf{Appendix: A general solution of steady momentum equations for thermocapillary
migration of a droplet at small Re numbers}

The momentum equations (7) of the continuous phase and the fluid in the droplet for the steady thermocapillary migration
of a droplet at small Re numbers are rewritten in terms of the stream functions $\Psi(r,\theta)$ and $\Psi'(r,\theta)$ as
\begin{equation}
\begin{array}{l}
\frac{1}{r^2 \sin \theta} (\frac{\partial \Psi}{\partial \theta}\frac{\partial}{\partial r}
-\frac{\partial \Psi}{\partial r}\frac{\partial}{\partial \theta} +2 {\rm ctg}\theta \frac{\partial \Psi}{\partial r}
-\frac{2}{r} \frac{\partial \Psi}{\partial \theta}) D^2 \Psi = \frac{1}{Re} D^4 \Psi,\\
\frac{\gamma}{r^2 \sin \theta} (\frac{\partial \Psi'}{\partial \theta}\frac{\partial}{\partial r}
-\frac{\partial \Psi'}{\partial r}\frac{\partial}{\partial \theta} +2 {\rm ctg}\theta \frac{\partial \Psi'}{\partial r}
-\frac{2}{r} \frac{\partial \Psi'}{\partial \theta}) D^2 \Psi' = \frac{\alpha}{Re} D^4 \Psi',\\
\end{array}
\end{equation}
where  $D^2 =\frac{\partial}{\partial r^2} +\frac{\sin \theta}{r^2}\frac{\partial}{\partial \theta}(\frac{1}{\sin \theta}\frac{\partial}{\partial \theta})$.
The boundary conditions at infinity and at the center of the droplet can be represented
in terms of the stream functions $\Psi(r,\theta)$ and $\Psi'(r,\theta)$ as
\begin{equation}
\begin{array}{l}
\Psi (r_\infty, \theta) \rightarrow -\frac{1}{2}V_\infty r^2_\infty \sin^2\theta,\\
  {\bf v}'(r_0, \theta) \sim \frac{\Psi'(r, \theta)}{r^2}|r_0 \sim O(1).
\end{array}
\end{equation}
In the same way, the boundary conditions (11) at the interface are rewritten as
\begin{equation}
\begin{array}{l}
v_r(R,\theta) = v_\theta(R,\theta){\rm tg} \phi,\\
v'_r(R,\theta) = v'_\theta(R,\theta){\rm tg} \phi,\\
 v_r(R,\theta){\rm tg} \phi + v_\theta(R,\theta) = v'_\tau(R,\theta){\rm tg} \phi+ v'_\theta(R,\theta),
\end{array}
\end{equation}
where $(v_r,v_\theta)=(\frac{1}{r^2\sin \theta}\frac{\partial \Psi}{\partial \theta}, -\frac{1}{r\sin \theta}\frac{\partial \Psi}{\partial r})$
, $(v'_r,v'_\theta)=(\frac{1}{r^2\sin \theta}\frac{\partial \Psi'}{\partial \theta}, -\frac{1}{r\sin \theta}\frac{\partial \Psi'}{\partial r})$
and $\phi(={\rm arctg} \frac{dR/d\theta}{R} )$ is an angle between ${\bf n}$ and ${\bf i_r}$.
Under the quasi-steady state assumption,
 the steady thermocapillary droplet migration requires that the
total net force acting on the droplet is zero. In particular, the zero net force  in the vertical direction is expressed as
\begin{equation}
 \int_{S_1} (\Pi_{r\theta} \sin \theta -\Pi_{rr} \cos \theta) dS =\int_0^{\pi} (\Pi_{r\theta} \sin \theta -\Pi_{rr} \cos \theta)|_R R^2 \sin \theta d\theta =0.
\end{equation}
In the momentum equations (51) with the boundary conditions (52)-(54), the controlled parameter is the Re number.
So the solutions of Eqs. (51) are sought in the forms of expansions in powers of the small parameter Re
\begin{equation}
\begin{array}{l}
\Psi(r,\theta) = \Psi^{(0)}(r,\theta) + Re \Psi^{(1)}(r,\theta) + o(Re),\\
\Psi'(r,\theta) = \Psi'^{(0)}(r,\theta) + Re \Psi'^{(1)}(r,\theta) + o(Re).
\end{array}
\end{equation}
Similarly, the pressure fields in Eqs. (7) have the perturbation expansions in the form
\begin{equation}
\begin{array}{l}
p(r,\theta) = \frac{1}{Re}p^{(0)}(r,\theta) + p^{(1)}(r,\theta) + o(1),\\
p'(r,\theta) = \frac{1}{Re}p'^{(0)}(r,\theta) + p'^{(1)}(r,\theta) + o(1)
\end{array}
\end{equation}
with the boundary conditions $p \rightarrow 0$  at infinity and $p'=p'_0=\frac{1}{Re} p'^{(0)}_0 + p'^{(1)}_0 + o(1)$ at the center of the droplet.
Substituting (55) and (56) into Eqs. (51)-(54), the governing equations and the boundary conditions for the zeroth order solutions
$\Psi^{(0)}(r,\theta)$ and $\Psi'^{(0)}(r,\theta)$ are, respectively, obtained as
\begin{equation}
\begin{array}{l}
D^4 \Psi^{(0)}(r,\theta)=0,\\
D^4 \Psi'^{(0)}(r,\theta)=0
\end{array}
\end{equation}
and
\begin{equation}
\begin{array}{l}
(i)\ \Psi^{(0)} (r_\infty, \theta) \rightarrow -\frac{1}{2}V_\infty r^2_\infty \sin^2\theta,\\
(ii)\ {\bf v}'^{(0)}(r_0, \theta) \sim \frac{\Psi'^{(0)}(r, \theta)}{r^2}|r_0 \sim O(1),\\
(iii)\ v^{(0)}_r(1,\theta) =0,\\
(iv)\ v'^{(0)}_r(1,\theta) =0,\\
(v)\ v^{(0)}_\theta(1,\theta) = v'^{(0)}_\theta(1,\theta),\\
(vi)\ \int_0^{\pi} (\Pi^{(0)}_{r\theta} \sin \theta -\Pi^{(0)}_{rr} \cos \theta)|_{1} \sin \theta d\theta =0.
\end{array}
\end{equation}
This is the classical model of the thermocapillary migration of a spherical droplet proposed by Young et al(1959)\cite{2}.
The solutions of Eqs. (57)  can be determined\cite{31,19} as
\begin{equation}
\begin{array}{l}
\Psi^{(0)}=-\frac{1}{2}V_\infty(r^2-\frac{1}{r}) \sin^2 \theta,\\
\Psi'^{(0)}=-\frac{3}{4}V_\infty r^2(r^2-1) \sin^2\theta,\\
p^{(0)}=0,\\
p'^{(0)}=p'^{(0)}_0 -15\alpha V_\infty r \cos \theta,
\end{array}
\end{equation}
where $p'^{(0)}_0$ is a constant,  $p^{(0)}(r,\theta)$ and $p'^{(0)}(r,\theta)$ are obtained
by integrating the zeroth order equations of Eqs. (7) with the boundary condition $p^{(0)}(r_\infty, \theta) \rightarrow 0$.
Meanwhile, the governing equations with the boundary conditions for the first order solutions
$\Psi^{(1)}(r,\theta)$ and $\Psi'^{(1)}(r,\theta)$ in Eqs. (51) are written as
\begin{equation}
\begin{array}{l}
D^4 \Psi^{(1)}(r,\theta)=\frac{1}{r^2 \sin \theta} (\frac{\partial \Psi^{(0)}}{\partial \theta}\frac{\partial}{\partial r}
-\frac{\partial \Psi^{(0)}}{\partial r}\frac{\partial}{\partial \theta} +2 {\rm ctg}\theta \frac{\partial \Psi^{(0)}}{\partial r}
-\frac{2}{r} \frac{\partial \Psi^{(0)}}{\partial \theta}) D^2 \Psi^{(0)}=0,\\
\alpha D^4 \Psi'^{(1)}(r,\theta)=\frac{\gamma}{r^2 \sin \theta} (\frac{\partial \Psi'^{(0)}}{\partial \theta}\frac{\partial}{\partial r}
-\frac{\partial \Psi'^{(0)}}{\partial r}\frac{\partial}{\partial \theta} +2 {\rm ctg}\theta \frac{\partial \Psi'^{(0)}}{\partial r}
-\frac{2}{r} \frac{\partial \Psi'^{(0)}}{\partial \theta}) D^2 \Psi'^{(0)}=0,
\end{array}
\end{equation}
and
\begin{equation}
\begin{array}{l}
(i)\ \Psi^{(1)} (r_\infty, \theta) \sim O(1),\\
(ii)\ {\bf v}'^{(1)}(r_0, \theta) \sim \frac{\Psi'^{(1)}(r, \theta)}{r^2}|r_0 \sim O(1),\\
(iii)\ v^{(1)}_r(1,\theta) =f'_1(\theta)v^{(0)}_\theta(1,\theta) -[v^{(0)}_r(R,\theta)-v^{(0)}_r(1,\theta)]/Re,\\
(iv)\ v'^{(1)}_r(1,\theta) =f'_1(\theta)v'^{(0)}_\theta(1,\theta) -[v'^{(0)}_r(R,\theta)-v'^{(0)}_r(1,\theta)]/Re,\\
(v)\ v^{(1)}_\theta(1,\theta) +[v^{(0)}_\theta(R,\theta)-v^{(0)}_\theta(1,\theta)]/Re= v'^{(0)}_\theta(1,\theta)+[v'^{(0)}_\theta(R,\theta)-v'^{(0)}_\theta(1,\theta)]/Re,\\
(vi)\ \int_0^{\pi} \{ (\Pi^{(1)}_{r\theta} \sin \theta -\Pi^{(1)}_{rr} \cos \theta)|_{1} + 2(\Pi^{(0)}_{r\theta} \sin \theta -\Pi^{(0)}_{rr}  \cos \theta)|_{1} f_1(\theta)\\
\ \ \ +[(\Pi^{(0)}_{r\theta}|_R-\Pi^{(0)}_{r\theta}|_{1}) \sin \theta -(\Pi^{(0)}_{rr}|_R -\Pi^{(0)}_{rr}|_{1} ) \cos \theta]/Re \} \sin \theta d\theta =0,
\end{array}
\end{equation}
where $D^2 \Psi^{(0)}=0$ and $D^2 \Psi'^{(0)}=-\frac{15}{2}V_\infty r^2 \cos^2 \theta$.
Using the expressions (59), the third condition at the interface in Eqs. (61) is expressed as
\begin{equation}
v^{(1)}_r(1,\theta) =3V_\infty [\frac{1}{2}f'_1(\theta) \sin \theta +f_1(\theta) \cos \theta]=\frac{1}{r^2 \sin \theta}\frac{\partial \Psi^{(1)}}{\partial \theta}|_1,
\end{equation}
which leads to $\Psi^{(1)} (r, \theta) \sim f_1(\theta) \sin^2 \theta $.
The function $f_1(\theta)$ is taken as $\omega \cos \theta$ ($\omega$ is a constant)
in order to  have $D^2 \Psi^{(1)} \sim f_1(\theta) \sin^2 \theta$.
Substituting a trial solution $\Psi^{(1)} (r, \theta)=F(r) \sin^2 \theta \cos \theta$  for the continuous phase into the first equation in Eqs. (60), we have
\begin{equation}
\Psi^{(1)}(r,\theta)=(\frac{A}{14}r^5+Br^3-\frac{C}{6}+\frac{D}{r^2})  \sin^2 \theta \cos \theta.\\
\end{equation}
Then, the first condition at infinity in Eq. (61) requires that
\begin{equation}
A=B=0.
\end{equation}
And the third condition at the interface in Eqs. (61), i.e., Eq. (62), demands that
\begin{equation}
D=\frac{3}{2} \omega V_\infty +\frac{C}{6}.
\end{equation}
Similarly, a trial solution of the form
\begin{equation}
\Psi'^{(1)} (r, \theta)= (\frac{A'}{14}r^5+B'r^3-\frac{C'}{6}+\frac{D'}{r^2})  \sin^2 \theta \cos \theta
\end{equation}
for the fluid in the droplet is suggested in terms of the fourth condition at the interface in Eqs. (61).
Then, the second condition at the center of the droplet in Eqs. (61) requires that
\begin{equation}
C'=D'=0.
\end{equation}
And the fourth condition at the interface in Eqs. (61) demands that
\begin{equation}
B'=\frac{3}{2} \omega V_\infty -\frac{A'}{14}.
\end{equation}
Using the expressions (63)(64) and the expressions (66)(67), the fifth condition at the interface in Eqs. (61) is reduced to
\begin{equation}
\frac{5}{14}A'+3B'+2C=\frac{15}{2} \omega V_\infty.
\end{equation}
Combining Eq. (68) and Eq. (69), we have
\begin{equation}
\begin{array}{l}
A'=14(\frac{3}{2}\omega V_\infty-C),\\
B'=C.
\end{array}
\end{equation}
Furthermore, the first order pressure fields  for the continuous phase and for the fluid in the droplet are determined by integrating the first order equations of Eqs. (7)
with the boundary condition $p^{(1)}(r_\infty,\theta) \rightarrow 0$
\begin{equation}
\begin{array}{l}
p^{(1)}(r,\theta)=-\frac{V^2_\infty}{4r^6} +[\frac{V^2_\infty}{2}(\frac{1}{r^3}-\frac{1}{4r^6})- \frac{C}{3r^3}](3 \cos^2 \theta -1),\\
p'^{(1)}(r,\theta)=p'^{(1)}_0 +\frac{3\gamma}{8}V^2_\infty r^4 +[\frac{3\gamma}{8}V^2_\infty (3r^2-2r^4)+\frac{\alpha A'}{2} r^2](3 \cos^2 \theta -1),
\end{array}
\end{equation}
where $p'_{01}$ is a constant.
Using the above stream functions $\Psi^{(0)}$, $\Psi^{(1)}$ and pressure fields $p^{(0)}$, $p^{(1)}$ for the continuous phase,
the sixth condition in Eqs. (61) is found to be self-perpetuating.

In addition to the above boundary conditions (53)(54) at the interface,
the deformed droplet in the steady thermopicallary migration has to satisfy
 other conditions based on the following facts. On the one hand, the volume
of the deformed droplet remains unchanged, i.e.,
\begin{equation}
\begin{array}{ll}
\Delta V &= \int_0^{\pi} \sin \theta d \theta (\int_0^R r^2 dr - \int_0^1 r^2 dr)\\
 &= \frac{1}{3}\int_0^{\pi}  (R^3 -1) \sin \theta d\theta\\
&\approx \frac{1}{3} \int^\pi_0 3 \omega Re \cos \theta \sin \theta d \theta +O(Re^2)\\
&\rightarrow 0.
\end{array}
\end{equation}
The condition is also self-perpetuating for the above choice of $f_1(\theta)=\omega \cos \theta$.
On the other hand, the center of mass of the droplet is always fixed at the origin of coordinates,
such as $\int_V z dV=0$ in the vertical direction.  It is further shown that
\begin{equation}
\begin{array}{ll}
\int_V z dV & =\int^\pi_0 \sin \theta d \theta \int^R_0 r \cos \theta r^2 dr\\
&=\frac{1}{4} \int^\pi_0 R^4 \sin \theta \cos \theta d \theta\\
%=\frac{1}{4} \int^\pi_0 (1 +2 \omega Re \cos \theta)^4 \sin \theta \cos \theta d \theta \\
&\approx \frac{1}{4} \int^\pi_0 [1 + 4 \omega Re \cos \theta +O(Re^2)] \sin \theta \cos \theta d \theta\\
&=\frac{2}{3} \omega Re +O(Re^2),\\
\end{array}
\end{equation}
which demands that $\omega=0$, i.e., $f_1(\theta)=0$. Finally, the first order stream functions for the continuous phase and for the fluid in the droplet are shown to be
\begin{equation}
\begin{array}{l}
\Psi^{(1)}(r,\theta)=-\frac{C}{6}(1-\frac{1}{r^2})  \sin^2 \theta \cos \theta,\\
\Psi'^{(1)} (r, \theta)= -\frac{C}{6} r^3(r^2-1)  \sin^2 \theta \cos \theta.
\end{array}
\end{equation}
The above derivations reveal that the functions in the solution of the steady momentum equations at small Re numbers
are independent of the temperature fields, but the parameters $V_\infty$, $C$, $p'^{(0)}_0$ and $p'^{(1)}_0$ in the solution
still depend on the temperature fields.

\end{document}